# Some New Aspects of Dendrimer Applications

Ophir Flomenbom[*], Roey J. Amir[++], Doron Shabat[++], and Joseph Klafter[*]

[*] **Department of Chemical Physics, School of Chemistry, Raymond and Beverly Sackler Faculty of Exact Sciences, Tel Aviv University**

[++] **Department of Organic Chemistry, School of Chemistry, Raymond and Beverly Sackler Faculty of Exact Sciences, Tel Aviv University**

**ABSTRACT**

Dendrimers are characterized by special features that make them promising candidates for many applications. Here we focus on two such applications: dendrimers as light harvesting antennae, and dendrimers as molecular amplifiers, which may serve as novel platforms for drug delivery. Both applications stem from the unique structure of dendrimers. We present a theoretical framework based on the master equation within which we describe these applications. The quantities of interest are the first passage time (FPT) probability density function (PDF), and its moments. We examine how the FPT PDF and its characteristics depend on the geometric and energetic structures of the dendrimeric system. In particular, we investigate the dependence of the FPT properties on the number of generations (dendrimer size), and the system bias. We present analytical expressions for the FPT PDF for very efficient dendrimeric antennae and for dendrimeric amplifiers. For these cases the mean first passage time scales linearly with the system length, and fluctuations around the mean first passage time are negligible for large systems. Relationships of the FPT to light harvesting process for other types of system-bias are discussed.

**Correspondence should be sent to**:

Ophir Flomenbom
Chemical Physics Department,
School of Chemistry,
Raymond & Beverly Sackler Faculty of Exact Sciences,
Tel Aviv University,
Ramat Aviv, Tel Aviv 69978, Israel
Tel: 972-640-7229
Fax: 972-640-6466
Email:  flomenbo@post.tau.ac.il

# 1. INTRODUCTION

Dendrimers are highly defined artificial macromolecules, which are characterized by a combination of a high number of functional groups, and a compact molecular structure [1]. The macromolecule constituents are organized in a branching form from a central core, creating a sphere of chemical end groups at the periphery that can be tailored according to the requirements [2, 3], see Fig. 1. The concept of repetitive-growth with branching can create unique spherical mono dispersed dendrimer formations, which are defined by their generation number [4]. As dendrimers are built from $\mathbf{AB}_z$-type monomers, each layer or generation ($G$) of branching unit doubles or triples, for $z=2$ or $z=3$, the number of peripheral groups. For example, for $z=2$, a first generation dendrimer, which is denoted *G1*, will have one branching unit, and a second generation dendrimer (*G2*) will have an additional two branching unit, etc. In addition the core branching $C$ can be chosen independently of $z$.

**FIGURE 1**

Although dendrimers are large molecules, they can be synthesized and characterized with a precision similar to that possible with smaller organic molecules. They do not suffer from the problem of 'poly-dispersity' that troubles linear macromolecules: that is constituents of a given set of dendrimers can have exactly the same molecular weight, rather than being a mixture of chains with a distribution of molecular weights [5]. The large number of identical chemical units in the branching sub-units, as well as those at the periphery confers greater versatility; see [6, 7] for examples.

The special features of dendrimers make them promising candidates for a large number of applications. For example, one can vary the type of the end groups, such that the specifically designed macromolecule can be used for sensing, catalysis or bio-chemical activity [8-18]. Most of the applications of dendrimers have been based mainly on the high number of functional groups. Although the enhanced effect that stems from many identical end groups being present simultaneously at the same place is of great importance, the combination of utilizing the multivalency and the precise architecture as an active framework able functioning more than just a scaffold, opens the way for novel and exciting dendrimeric devices. Two such concepts that conceal both the multivalency and the active framework features are the use of dendrimers as light harvesting antennae [19-22], and as molecular amplifiers [7, 21, 23, 24]. The latter use of dendrimers is based on disintegrating the dendrimeric skeleton into its subunits, thus releasing all of its functional end-groups as a result of a signal initiated at the core of the dendrimer. The amplification is established due to the exponential increase of the number of subunits along the dendrimer's generations (Fig. 2). It has been suggested that when drug molecules are attached as end groups at the periphery, the dendrimer can be used as an efficient drug delivery platform [25]. For latest developments in this direction the reader is referred to Refs. [7, 21, 23-25].

**FIGRUE 2**

When using dendrimers as antennae, one utilizes the active framework as an energy funnel, which directs excitation energy from donors placed at the periphery of the dendrimer to its core, where an acceptor molecule is placed (Fig. 3). In this work we focus on this dendrimeric application. We present a model that describes the migration process of the excitations across the dendrimeric framework, and discuss the properties of such possible antennae.

**FIGURE 3**

## 2. DENDRIMERS AS LIGHT HARVESTING ANTENNAE

In recent years efforts have been devoted to create and manipulate molecular scale systems that can be used as efficient light harvesting antennae [15-20, 22]. Dendrimers, due their special architecture, have been proposed as candidates for serving as antennae. Yet, an issue for debate is whether the energy from the periphery, where chromophores are placed, is being transferred through space directly to the dendrimeric core, or alternatively, through the bonds of the dendrimeric framework, thus using the special structure of the molecule, as is shown in Fig. 3. The main argument for a direct energy transfer relies on the fact that most dendrimers are synthesized through *meta*-position branching that prevents resonative conjugation among the benzene rings, and leads to localization of the $\pi$-electron excitations [22]. However, it has been demonstrated that by designing dendrimers with varying generation lengths, one can create an energetic funnel which is directed towards the dendrimeric core, in particular, when the length between generations decreases for distant core generations [22]. Such dendrimers are termed extended, in contrast to compact dendrimers for which the length between generations is fixed. More recently dendrimers that are branched through *meta*-position and *para*-position were synthesized [26], thus opening another possibility for building dendrimeric antenna. In addition to the demand that the dendrimer is design such that energetic funnel is created, the demand for a rigid dendrimeric structure is required for energy transfer through bonds. A rigid structure helps energy transfer through bonds competing favorably against direct energy transfer.

Here, we present a theory that models a dendrimeric antenna assuming that the excitation energy migrates along the bonds. In subsection 2.1. we formulate the model, which is then studied in details in subsections 2.2. through 2.4. The presented model is general and can be applied for other systems as well [27, 28]. Nevertheless throughout this section relationships to dendrimers are emphasized, and in particular subsection 2.5 is dedicated to the thermodynamic aspects of dendrimeric antennae.

### 2.1. Formulation of the model

The dynamics of an excitation (signal) spreading over a dendrimer can be described as a one-dimensional nearest neighbors hopping process [19, 20]. This mapping is valid if one is interested only in the distance of the excitation from the core, and not on its exact position within a generation. Thus, the time evolution of the signal is given by a set of coupled kinetic equations with an absorbing and reflecting boundary conditions:

$$\frac{\partial}{\partial t} P_0(t) = a_1^f P_1(t); \qquad \text{Absorbing site,} \qquad (1a)$$

$$\frac{\partial}{\partial t} P_1(t) = -(a_1^f + a_1^b) \cdot P_1(t) + a_2^f \cdot P_2(t), \qquad (1b)$$

$$\frac{\partial}{\partial t}P_m(t) = a_{m-1}^b P_{m-1}(t) - (a_m^f + a_m^b) \cdot P_m(t) + a_{m+1}^f \cdot P_{m+1}(t); \quad 2 \leq m \leq n-1, \tag{1c}$$

$$\frac{\partial}{\partial t}P_n(t) = a_{n-1}^b \cdot P_{n-1}(t) - a_n^f \cdot P_n(t), \qquad \text{Reflecting site.} \tag{1d}$$

We introduce a reflecting boundary at the periphery (site $j = n$; see Fig. 4) assuming that the signal stays in the system as long as it does not reach the core, i. e. the absorbing site $j = 0$. Once the signal reaches the absorbing site it is captured there. The other symbols appear in Eqs. (1a)-(1d) are: $P_j(t)$ is the probability density function (PDF) that the signal occupies site $j$ at time $t$, and $a_j^f$ ($a_j^b$) is the transition rate from site $j$ to site $j$-1 ($j$+1). Figure 4 presents a schematic illustration of the system. The coupled equations (1a)-(1d) are also referred to as the system's master equation. For the case of a dendrimeric amplifier, where the signal propagations along the dendrimer results in its irreversible disassociation, all the backward rates in Eqs. (1a)-(1d) should be set to zero. This results in a simplified set of equations,

$$\frac{\partial}{\partial t}P_0(t) = a_1^f P_1(t); \qquad \text{Absorbing site,} \tag{1e}$$

$$\frac{\partial}{\partial t}P_m(t) = -a_m^f \cdot P_m(t) + a_{m+1}^f \cdot P_{m+1}(t); \qquad 1 \leq m \leq n-1, \tag{1f}$$

$$\frac{\partial}{\partial t}P_n(t) = -a_n^f \cdot P_n(t); \qquad \text{Reflecting site.} \tag{1g}$$

**FIGURE 4**

The site index $j$ represents the $j^{th}$ generation of the dendrimer. The dependence of the transition rates on $j$ should reflect both the exponential branching of the dendrimer end groups with the increase in the generation number, which can be viewed as an entropic bias towards the periphery [19, 20], and a possible energetic funnel towards the core. We elaborate on this issue in subsection 2.5.

It is convenient to write the master equation expressed by equations (1a)-(1d) in a matrix representation,

$$\partial \vec{P}(t)/\partial t = \mathbf{A}\vec{P}(t), \tag{2}$$

where $[\vec{P}(t)]_j = P_j(t)$ is the occupation PDF of site $j$. The propagation matrix $\mathbf{A}$ is a tridiagonal $n$-dimensional square matrix (the absorbing site is not included in the propagation matrix) that contains information about the time evolution of the signal, in term of the transition rates, $a_j^f$'s and $a_j^b$'s.

The formal solution of the master equation is

$$\vec{P}(t) = \mathbf{C}\exp(\lambda t)\mathbf{C}^{-1}\vec{P}_0, \tag{3}$$

where $[\vec{P}_0]_j = \delta_{j,x}$ is the initial condition (the process starts in site $x$ with probability one), $\boldsymbol{\lambda}$ is the eigenvalue matrix obtained through the similarity transformation $\boldsymbol{\lambda} = \mathbf{C}^{-1}\mathbf{A}\mathbf{C}$, and $\mathbf{C}$ and $\mathbf{C}^{-1}$ are, respectively, the eigenvector matrix and its inverse of $\boldsymbol{\lambda}$.

Due to the normalization condition we have $P_0(t) + \sum_{j=1}^{n} P_j(t) = 1$, where $P_0(t)$ is the PDF to occupy the trap at time $t$. We further define $S(t) = \vec{U}_n \vec{P}(t) = \sum_{j=1}^{n} P_j(t)$ as the survival probability which is the probability that the signal has not been absorbed, where $\vec{U}_n = (1,1,:,1)$ is the summation row vector on $n$ dimensions. Note that $P_0(t)$ increases monotonically with time, while $S(t)$ decreases monotonically with time. Both functions depend on the initial condition $x$. Accordingly, from this point on, we add the initial condition $x$ as an additional variable in the survival probability, $S(t,x)$, and functions that are derived from it. A function that plays a central role in the theory of random walks in finite and semifinite systems, and is of great importance in applications, is the first passage time (FPT) PDF $\Phi(t,x)$ [29]. $\Phi(t,x)$ is defined by

$$\Phi(t,x) = \frac{\partial}{\partial t}[1 - S(t,x)], \tag{4}$$

which is equivalent to Eq. (1a); namely, the FPT PDF is the rate of change of the trap occupation probability. By comparing the temporal behavior of Eq. (4) to experimental results, one can extract the system transition rates that appear in Eqs. (1a) - (1d). We note that in a recent work [30], the FPT PDF was studied and related to dendrimeric antennae by a similar approach, although for the case where matrix $\mathbf{A}$ represents only the branching of the dendrimer. The importance of computing the PDF of the FPT and not just its moments is emphasized, for example, by single molecule experiments where the PDF is measured directly. Experiments involving individual dendrimers have been reported recently [31].

**2.2. The first passage times PDF**

Below we show some properties of the FPT PDF in Eq. (4) for several invariant systems, which are of interest to dendrimers. By an invariant system we mean that the transition rates are taken to be independent of the site (generation) index $j$, namely, $a_j^b = k_-$ and $a_j^f = k_+$ for all $j$. For these systems the ratio between the forward and the backward transition rates, $Q \equiv k_- / k_+$, is the relevant parameter. We distinguish between three cases, $Q \ll 1$, $Q = 1$ and $Q \gg 1$:
(1) The first case $Q \ll 1$ represents a system that displays a bias towards the absorbing site. Such a choice can describe a dendrimer that has a large energetic bias towards its core [19, 20, 32], or a dendrimer that irreversibly dissociates as the signal propagates across the molecule as in molecular amplifiers [7, 23-25], and multi-triggering self-immolative dendrimers [33].
(2) The second case $Q = 1$ describes bias-free dynamics, which means that the entropic bias is canceled by the energetic bias in the corresponding dendrimeric antennae systems.

(3) The third case, $Q \gg 1$, represents an escape process against a constant force, which for our purposes is translated into a dendrimeric antenna that experiences an energetic bias towards the periphery.

Figure 5a shows $\Phi(\tau,n)$ for an invariant system and $Q = 0.01$, as a function of the dimensionless time $\tau \equiv tk_+$ for several system sizes, $n = 2, 3, 4, 6$, and $x = n$, namely, the process starts at the reflecting site (the periphery). $\Phi(\tau,n)$ is mono-peaked for all $n$ values (and in general for all $n>1$) and decays exponentially at large times. Figure 5b shows three characteristics of a density function, here for $\Phi(\tau,6)$. These are the average $<\tau>$, the standard deviation $\sigma = \sqrt{<\tau^2> - <\tau>^2}$, and their ratio $R = \sigma/<\tau>$, also known as the relative error of the PDF. The relative error is an important characteristic that gives a normalized measure for a spread of a density function. We discuss this quantity further in subsection 2.4. In Figs. 5c - 5d we show the dependence of these characteristics on $n$. Note that $<\tau(n)> \sim n$, and $R(n) \sim \sqrt{1/n}$, for $Q \ll 1$.

**FIGURE 5**

Figure 6a compares $\Phi(\tau,4)$ of an invariant *symmetric* system, namely $Q = 1$, to a trap-oriented ($Q \ll 1$) system with $Q = 0.01$. Although the location of the peak of the PDFs is similar for both cases, $\Phi(\tau,4)$ for the symmetric case is broader. This is also reflected in the slower saturation of the cumulative probability function of the first passage times PDF $G(\tau,n) = \int_0^\tau \Phi(s,n)ds$, shown in the inset of Fig. 6a for $n = 4$. For some applications, $G(\tau,n)$ is the direct information obtained from experiments [34], and can be used to obtain the mean of $\Phi(\tau,n)$. For the symmetric case the mean of the PDF scales as $n^2$ (Fig. 6b), while the relative error $R$ is independent of $n$ for large systems (inset of Fig. 6b), in contrast to the trap-biased system.

**FIGURE 6**

For the third case of an invariant system with a bias towards the periphery ($Q = 2.5$), $\Phi(\tau,4)$ decays slowly relative to its symmetric invariant counterpart (Fig. 7a). Note that as $Q$ increases the smallest absolute eigenvalue, $|\lambda_{min}|$, dominates the PDF behavior, which leads to the relation $<\tau> \approx 1/|\lambda_{min}|$. Figure 7b shows that for this case the mean of the PDF grows exponentially with the system size, $<\tau> \propto e^{n \log Q}$, and the relative error is approximately unity (inset).

**FIGURE 7**

## 2.3. The mean first passage time

The characteristics of a PDF are its moments and their interrelations. The *m>0* moment of the first passage times PDF is defined by:

$$<t^m(x)> = \int_0^\infty t^m \Phi(t,x)dt = m\int_0^\infty t^{m-1} S(t,x)dt. \quad (5)$$

Here, the second equality is obtained when integrating by parts, using Eq. (4), and noticing that the boundary term vanishes due to the fact that the survival probability is zero at infinite times. Substituting $m=1$ in Eq. (5), the mean first passage time (MFPT) is obtained:

$$<t(x)> = \int_0^\infty S(t,x)dt = -\vec{U}_n \mathbf{A}^{-1} \vec{P}_0, \quad (6)$$

where the second equality is obtained by using Eq. (3). Thus, one needs to invert the matrix $\mathbf{A}$ to calculate the MFPT. Note that as long as all the forward transition rates are finite, matrix $\mathbf{A}^{-1}$ exists. Equation (6) can be written as

$$<t(x)> = \sum_{j=1}^n t_{j,x}; \qquad t_{j,x} \equiv -\mathbf{A}^{-1}_{j,x}, \quad (7)$$

where $t_{j,x}$ is the mean residence time (MRT) of site $j$ when starting at site $x$, before trapping occurs [35]. Although $t_{j,x}$ can be expressed in terms of the transition rates for any system size and arbitrary choice of the transition rates [35, 36], we present below the MFPT for several invariant systems, namely, $a_j^b = k_-$ and $a_j^f = k_+$ for all generations $j$. For these cases, the MFPT reads [35]:

$$<t(x)> = \frac{1}{\Delta k}[x + \frac{Q^{n+1}}{1-Q}(1-Q^{-x})] \quad \text{for} \quad Q \equiv \frac{k_-}{k_+} \neq 1 \quad (8)$$

$$<t(x)> = \frac{x(2n+1-x)}{2k} \quad \text{for} \quad k_- = k_+ \equiv k \quad (9)$$

where $\Delta k = k_+ - k_-$.

For large biases, $Q \ll 1$ and $Q \gg 1$, Eq. (8) reduces to

$$<t(x)> = \begin{cases} x/\Delta k & ; \quad Q \ll 1 \\ e^{(n+1)\ln Q}(1-Q^{-x})/[(1-Q)\Delta k]; & Q \gg 1 \end{cases} \quad (10)$$

Equations (9) and (10) demonstrate the dependence of the MFPT on the system size. For a system that displays a bias towards the center of the dendrimer, the average time to be trapped scales linearly with the initial site of the process, $x$. For a system which is biased towards the periphery, an exponential dependence of the MFPT with the system size is exposed, regardless of the initial site of the process. For a system with no bias at all, namely $k_- = k_+$, the scaling of the MFPT with the size of the system depends on the initial site; when $x=1$ a linear scaling with the system size is evident, while a square scaling with the system size is obtained for $x=n$.

For the special case of an invariant system and $k_- \to 0$ (we term this system a "death" system) a general expression for the $m^{th}$ moment is valid. For starting at the reflecting site $x=n$, we have:

$$<t^m(n)> = \frac{1}{k^m} \frac{(n+m-1)!}{(n-1)!} . \qquad (11)$$

Equation (11) gives a full characterization of first passage time PDF of a "death" process, and can be used to obtain the first passage time PDF by inverting the Laplace transform of $\Phi(t)$, $\overline{\Phi}(s) = \int_0^\infty \Phi(t) e^{-st} dt = \sum_m <t^m> (-s)^m / m!$. This procedure yields

$$\Phi(t) = k(kt)^{n-1} e^{-kt} /(n-1)! . \qquad (12a)$$

Equation (12a) is the Poisson PDF. This corresponds to very efficient antennae, and to the dendrimeric amplifier (Fig. 3). In the latter case, the signal propagation along the dendrimer results in its irreversible disassociation. For a situation where each of the single events occurs with a specific rate, the solution of $\Phi(t)$, which is obtained from solving Eqs. (1e)-(1g), reads

$$\Phi(t) = \sum_{i=1}^n a_i^f e^{-a_i^f t} A_i \qquad ; \qquad A_i = \prod_{j \neq i} a_j^f /(a_j^f - a_i^f) . \qquad (12b)$$

Note that the results in this subsection give a measure for the efficiency of dendrimeric antennae when assuming that the excitation energy transfers through bonds and that multiple excitation do not play a role. However for multi-excitation system the MFPT of the first excitation to reach the core is shorter than the MFPT presented here, and depends on the number of excitation the process starts with [37]. For the case of the dendrimeric amplifier, the quantity that measures the efficiency of the amplification should involve the exponential branching factor, which represents the number of released molecules, in addition to the temporal behavior given by Eq. (12b).

**2.4. The second moment of the first passage times PDF**

The second moment of a PDF provides information about its spread. For our model, the expression for the second moment $<t^2(x)>$ reads

$$<t^2(x)> = 2\vec{U}_n \mathbf{A}^{-1} \mathbf{A}^{-1} \vec{P}_0 , \qquad (13)$$

which is obtained by using Eqs. (4) and (5). Rewriting Eq. (13) as

$$<t^2(x)> = 2\sum_{j=1}^n <t(j)> t_{j,x} , \qquad (14)$$

the second moment can be calculated for an arbitrary choice of the transition rates. As mentioned, here we are interested in invariant systems. By straightforward calculations we obtain the second moment for an invariant system and for $Q \neq 1$,

$$<t^2(n)> = \left(\frac{\sqrt{2}}{\Delta a}\right)^2 \left[\frac{n(n+1)}{2} + \frac{Q^{n+1}(2n+1) - Q^{n+2}(3n-1) + Q^{2n+2} - 2Q}{(1-Q)^2}\right], \qquad (15)$$

which for large *n* systems reduces to,

$$<t^2(n)> = \begin{cases} (n+1)n/2 & ; \quad Q \ll 1 \\ 2Q^{2n+2}/[\Delta a(1-Q)]^2 & ; \quad Q \gg 1. \end{cases} \qquad (16)$$

For the symmetric invariant system $Q = 1$, we get

$$<t^2(n)> = \frac{n(n+1)}{12k^2}(5n^2 - 5n + 2). \qquad (17)$$

The relative error introduced in subsection 2.1. is an essential parameter in the statistics here. The relative error is the ratio between the standard deviation, $\sigma$, and the average of a density function both depending on the initial site, $R(x) = \sigma(x)/<t(x)>$, where $\sigma(x) = \sqrt{<t^2(x)> - <t(x)>^2}$. This ratio provides a measure for the spread of a density function regardless of its argument values. For a sufficiently large system, we get for invariant systems and $x = n$:

$$R(n) = \begin{cases} \sqrt{1/n} & ; \quad Q \ll 1 \\ \sqrt{2/3} & ; \quad Q = 1 \\ 1 & ; \quad Q \gg 1. \end{cases} \qquad (18)$$

Equation (18) can be compared with the numerical results (Figs. 5 - 7). Note that only a trap-oriented system exhibits the desirable behavior of $R(n)$, namely, $R(n)$ vanishes (as $1/\sqrt{n}$) for large systems.

## 2.5. Thermodynamics and dendrimers

In this subsection we describe thermodynamically a dendrimeric antenna [38], from which we derive an explicit expression of one of the transition rates for an invariant systems. We consider an excitation that migrates on a dendrimer by nearest neighbors jumps. The core is assumed to capture the energy for some time, depending on its release rate. Namely, the core functions as a reversible trap (later on we take the limit of an irreversible trap). Energy levels are assigned for each generation, such that a funnel towards the core is created

$$\begin{aligned} E_0 &= \varepsilon_0 \\ E_G &= \varepsilon_0 + \varepsilon_1 + (G-1)U \; ; \; 1 < G \leq n, \end{aligned} \qquad (19)$$

where $\varepsilon_0$ is the core excitation energy, $\varepsilon_1$ is the excitation energy difference between the core excitation energy and the first generation excitation energy, and $U$ is the excitation energy difference between each nearest neighbors generations (Fig. 8). The excitation energy levels descend from the periphery to the core, which creates an energetic funnel (Fig. 9).

**FIGURE 8**

The structure of a dendrimer with branching $z$ towards the periphery, and a core branching $C$, leads to the degeneracy

$$f_G = Cz^{G-1}. \tag{20}$$

Having assigned excitation energies and degeneracy to each of the generations, the partition function of the dendrimer is

$$Z = e^{-\varepsilon\beta} + \sum_{G=1}^{n} f_G e^{-E_G\beta} = e^{-\varepsilon\beta} + Ce^{-(\varepsilon+\varepsilon_1)\beta} \sum_{G=1}^{n} [ze^{-U\beta}]^{G-1}, \tag{21}$$

where $\beta^{-1} = k_B T$, $k_B$ is the Boltzmann constant and $T$ is the temperature. The equilibrium occupation probabilities of the various generations are

$$\begin{aligned} P_{0,eq} &= e^{-\varepsilon\beta}/Z \\ P_{G,eq} &= P_{0,eq} Ce^{-\varepsilon_1\beta}[ze^{-U\beta}]^{G-1} \end{aligned} \tag{22}$$

from which the free energy of each generation follows,

$$F_G = -\beta^{-1}\ln(P_{G,eq}Q) = E_G - \beta^{-1}\ln(f_G). \tag{23}$$

Fig. 10 shows the free energy as a function of the generation index for three values of the parameter $\Delta = \ln(z)/(U\beta)$, $\Delta = 0, 1, 2$. For $\Delta = 0$ an energetic funnel towards the the trap, similar to that shown in Fig. 9, is created. For $\Delta = 1$ there is no energetic preference to be at a specific generation (excluding the trap), whereas for $\Delta = 2$ an energetic preference towards the periphery exists. Namely at high temperatures the energetic funnel becomes less efficient relative to the geometric one.

**FIGURE 9**

**FIGRUE 10**

Now, we wish to translate the energetic picture into the dynamical model by obtaining expressions for the transition rates from the thermodynamic picture. To do so, we used the equilibrium condition [38, 39]

$$P_{eq,G} a_G^b = P_{eq,G+1} a_{G+1}^f. \tag{24}$$

From Eqs. (22) and (24) it follows that the transition rates are given by

$$k_- = k_+ z e^{-U\beta}$$
$$a_0^b = k_+ C e^{-\varepsilon_1 \beta} \qquad (25)$$

and $k_+$ is arbitrary. Note that $k_-$ has two components; the component $z$ of geometric origin, whereas $e^{-U\beta}$ emerges from energetic consideration.

Taking the limit $\varepsilon_1 \beta \gg 1$ in Eq. (22), leads to $a_0^b = 0$, we thus recover the description of an irreversible trap shown in Fig. 4. Accordingly, the expression for the moments of the first passage times PDF, Eqs. (9) and (10) and Eqs. (16) and (17), can be written now in terms of the energetic model described in this subsection.

## 3. Concluding Remarks

Dendrimeric applications exploit the special architecture of dendrimers to build nano-devices. Both periphery-to-core and core-to-periphery processes are possible in applications. This is demonstrated by using dendrimers as antennae and as amplifiers. In this work, we have studied mainly the use of dendrimers as efficient light harvesting antennae. We have investigated the properties of the first passage time PDF of a signal migrating on the dendrimer to reach the core from the periphery. Calculations of the first two moments of this PDF as a function of the system size *n*, and the parameter *Q* have been given for invariant systems. The parameter *Q* determines the process-bias, namely for large (small) values of *Q* there is a bias towards the periphery (core). It has been shown that the dependence of the mean of the first passage time PDF on *n* changes from exponential to quadratic and then to linear, for periphery-oriented, non-biased, and core-oriented systems, respectively. The fluctuations around the mean have been shown to scale with *n* as the scaling of the mean for the first two systems, and as the square root of the mean for the core-oriented system. It is clear that when designing efficient antennae, one should try to build a device which has a bias towards the region that collects the light. The first two moments of the core-oriented system characterizes the efficiency of such antennae. Moreover, in the limit of a very large bias ($Q \to 0$), the exact solution of the first passage time PDF can be obtained, and is simply the Poisson PDF. For a more general system, where each of the single events occurs with a different rate, the FPT PDF, which is given by Eq. (12b), is a sum of weighted exponentials whose rates being the rates of the single events. This case of very large bias is suitable to describe dendrimeric amplifiers and multi-triggering self-immolative dendrimers as well.

Finally, by using two simple physical arguments, which are the entropic-like bias of the dendrimeric exponential branching, and energetic levels assigned to each generation in the dendrimer as a function of its distance from the core, relationships between the system rate constants that determine *Q* have been proposed.

**Figure Caption:**

FIG 1 Schematic representation of a dendrimer. Shown are the generation indices, the dendrimer core and end groups. Here $z=2$.

FIG 2 Dendrimer as an amplifier. Upon trigger activation the dendrimer framework undergoes a spontaneous disassembly to release its end groups.

FIG 3 Schematic illustration of a dendrimer as an antenna. The excitation energy $h\nu$ (shown in red) migrates along the denrimeric framework until it reaches the core where the energy is used for some purpose such as a chemical reaction.

FIG 4  **a -** Schematic illustration of the escape process from a system with $n$ generations (sites). Once the signal reaches site $j=0$ it is absorbed, namely site $j=0$ is a trap. **b -** The mapping of a two-generation dendrimer onto a one-dimensional system. See text for the corresponding discussion.

FIG 5 **a -** $\Phi(\tau,n)$ for $Q=0.01$, and several values of $n$, $n=2, 3, 4$, and $6$, correspond to the full, dotted-dashed, dashed, and dotted lines, respectively. **b -** The parameters $R$, $\sigma$, and $\tau$, are shown for $\Phi(\tau,6)$. **c -** The mean of the density functions displays a linear scaling with $n$, and **d** – the relative error $R$ decays as $n^{-1/2}$.

FIG 6 **a -** $\Phi(\tau,4)$ and $G(\tau,4)$ (inset) for two values of $Q$, $Q=0.01$ (full curves), and $Q=1$ (dashed curves). **b -** The mean for the symmetric case scales as $n^2$, and $R$ reaches its asymptotic value $[=(2/3)^{1/2}$, see Eq. (18)] at considerably small systems (inset).

FIG 7 **a** – The slowly decaying $\Phi(\tau,4)$ for $Q = 2.5$ in comparison to the symmetric case, and **b** – The characteristics of the PDF for $Q=50$ are $<\tau(n)> \approx e^{n\log Q}$, and $R \approx 1$ (inset).

FIG 8 A schematic illustration of a four generation dendrimer with $z=2$ and $C=2$. Also shown are the energies of each generation $E_4>E_3>E_2>E_1$.

FIG 9 The excitation energy as a function of the generation index. Here, $\varepsilon=2$, $\varepsilon_1=6$, $U=1$, and $n=9$. The first point corresponds to the core (trap).

FIG 10 The free energy as a function of the generation number, for three different values of $\Delta$, $\Delta=0, 1, 2$. Here $z=2$ and $C=2$.

Fig 1

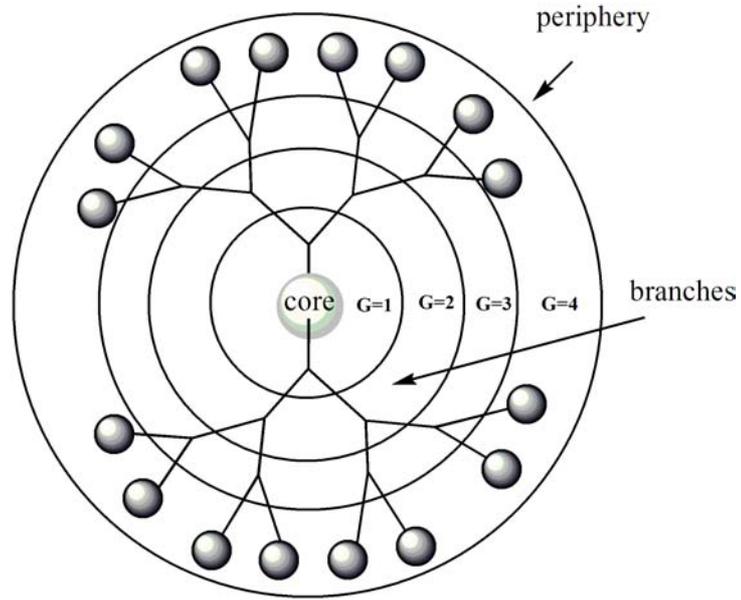

FIG 2

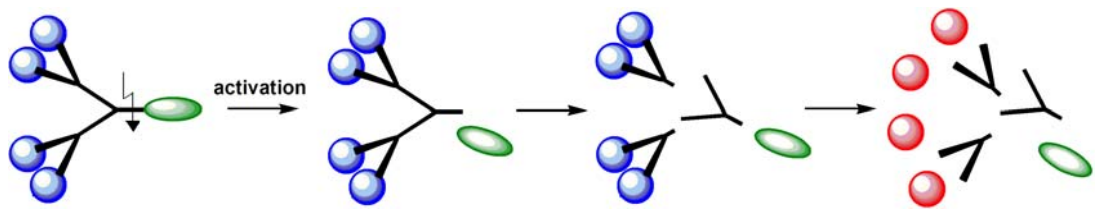

FIGURE 3

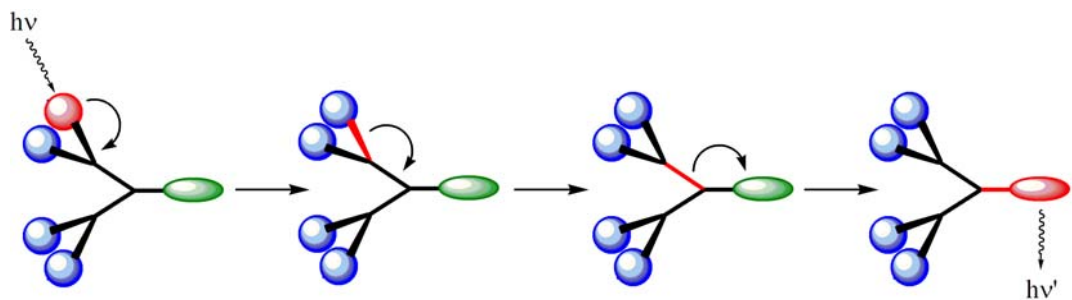

FIGURE 4

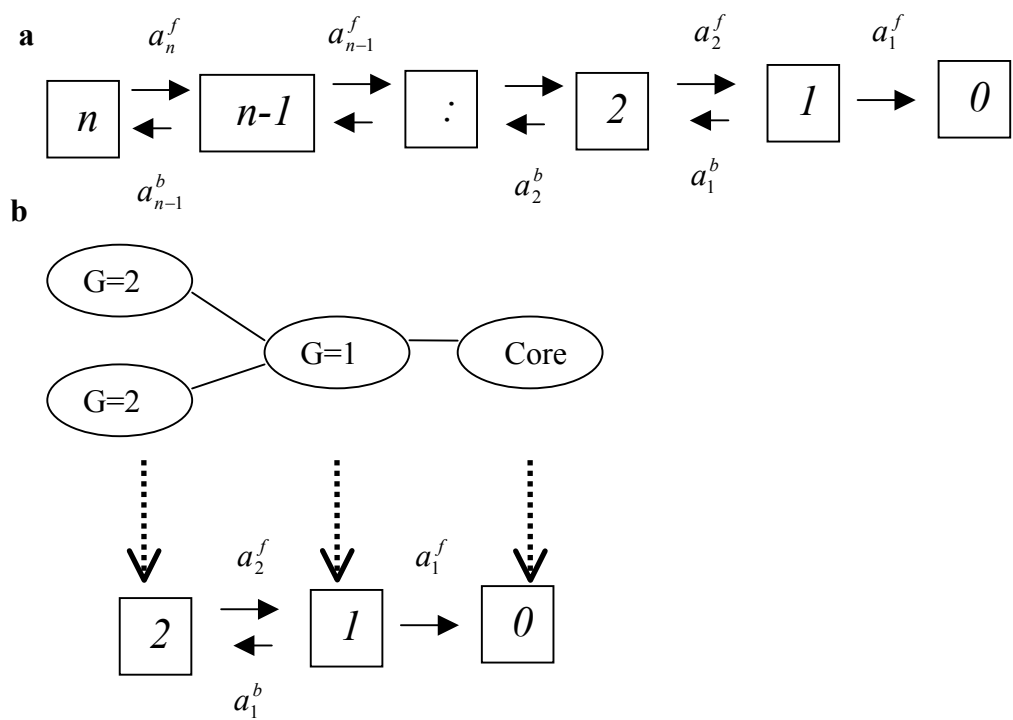

FIGURE 5

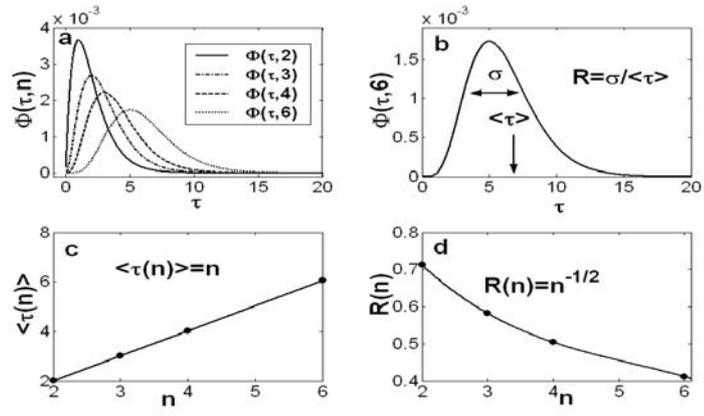

FIGURE 6

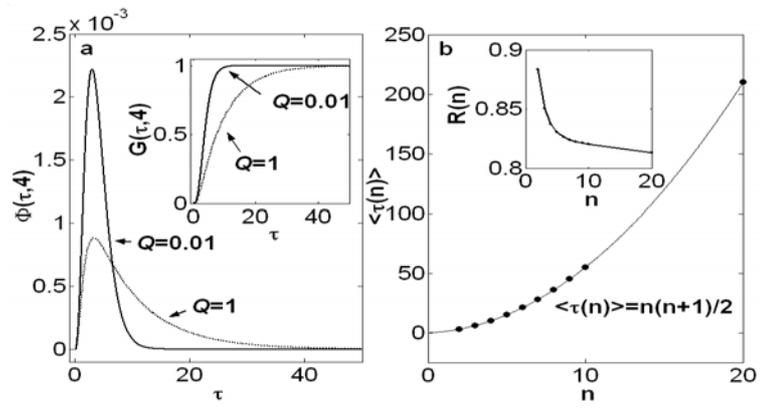

FIGURE 7

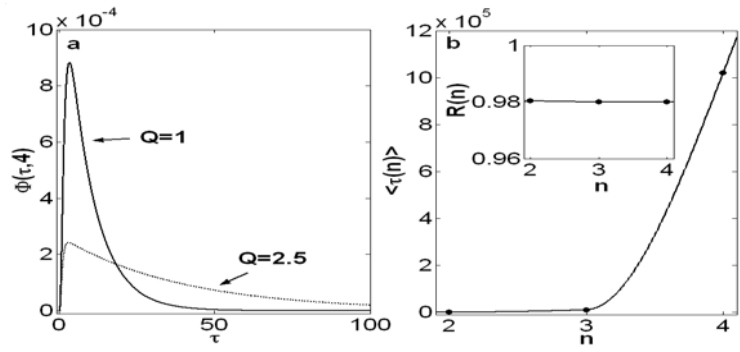

FIGURE 8

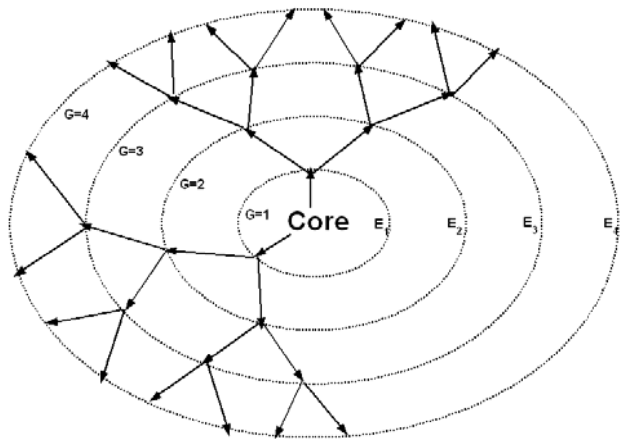

FIGURE 9

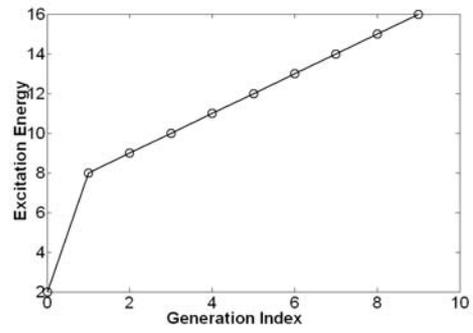

FIGURE 10

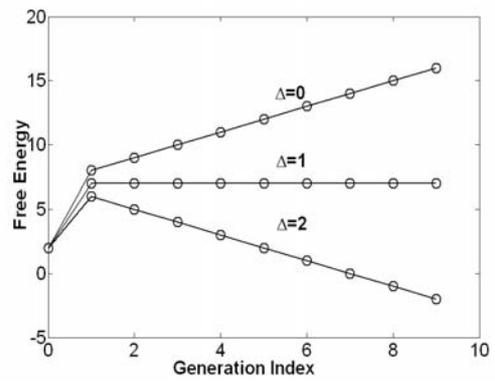